\documentclass[12pt]{article}

\usepackage{epsfig}
\include{amssym}

\def\mathbf{\vec}

\def\ca{\c{c}\~{a}}

\newcommand{\njl}{\mathrm{NJL}}

\newcommand{\A}{\mathrm{A}}
\newcommand{\h}{\mathrm{H}}

\newcommand{\st}{\mathrm{st}}

\newcommand{\eff}{\mathrm{eff}}
\newcommand{\ud}{\mathrm{d}}

\begin{document}

\centerline{\Large\bf\sc Multi-quark interactions with a globally} 

\vspace{0.2cm}
\centerline{\Large\bf\sc stable vacuum} 
\vspace{1cm}

\centerline{\large A. A. Osipov\footnote{On leave from Joint 
            Institute for Nuclear Research, Laboratory of Nuclear 
            Problems, 141980 Dubna, Moscow Region, Russia. 
            Email address: osipov@nusun.jinr.ru}, 
            B. Hiller\footnote{Email address: 
            brigitte@teor.fis.uc.pt},
            J. da Provid\^encia\footnote{Email address:
            providencia@teor.fis.uc.pt}} 
\vspace{0.5cm}
\centerline{\small\it Centro de F\'{\i}sica Te\'{o}rica, 
            Departamento de
            F\'{\i}sica da Universidade de Coimbra,}
\centerline{\small\it 3004-516 Coimbra, Portugal}
\vspace{1.5cm}

\centerline{\bf Abstract}
\vspace{0.5cm}
It is shown that $U(3)_L\times U(3)_R$ eight-quark interactions 
stabilize the asymmetric ground state of the well-known model with 
four-quark Nambu -- Jona-Lasinio and six-quark 't Hooft interactions. 
The result remains when the reduced $SU(3)$ f\mbox{}lavour symmetry is 
explicitly broken by the general current quark mass term  
with $\hat{m}_u\neq\hat{m}_d\neq\hat{m}_s$.

\vspace{7.0cm}
PACS number(s): 12.39.Fe, 11.30.Rd, 11.30.Qc

\newpage 

\section*{\centerline{\large\bf 1. Introduction}}

Phenomenological parametrizations based on some simple ansatz with 
solid symmetry grounds are frequently used in low-energy QCD. One of 
the most common and important outputs of this approach is to get a
clue of how high energy QCD may inf\mbox{}luence low energy observables. 
Unfortunately, in spite of all remarkable successes of the QCD sum 
rules method \cite{Shifman:1979}, or chiral perturbation theory 
\cite{Gasser:1984}, this picture is still far away from being
completed. 
 
Some features of the large distance hadron dynamics can be understood 
in the framework of ef\mbox{}fective chiral Lagrangians written in
terms of quark degrees of freedom \cite{Meissner:1988}. They are 
ef\mbox{}ficient for the description of spontaneous chiral symmetry 
breaking, or for the study of the quark structure of light mesons. The 
parameters of such Lagrangians can be related to the characteristics
of the QCD vacuum given in form of the vacuum expectation values of
the relevant quark bilinears or gluons (if they are included). In many 
respects this approach corresponds to a Landau -- Ginzburg like 
description of the f\mbox{}lavour dynamics. 

There are a number of instructive models that assume the existence of 
underlying multi-quark interactions and their importance for physics
of hadrons. Well-known examples include the Nambu and Jona-Lasinio 
(NJL) model \cite{Nambu:1961}, where the four-fermion interactions 
have been used to study dynamical breaking of chiral 
symmetry\footnote{Later on a modified form of this interaction has 
been used to derive the QCD ef\mbox{}fective action at long distances 
\cite{Eguchi:1976,Volkov:1982,Ebert:1986}.}; the instanton inspired 
models \cite{Diakonov:2003}, where $2N_f$-quark interactions 
($N_f$ is the number of quark f\mbox{}lavours) offer a possible 
framework to discuss the $U_\A (1)$ problem \cite{Hooft:1978}; the
potential-type quark models which are successfully applied to the 
evaluation of hadronic parameters \cite{Isgur:1985}. 
   
In this Letter, we propose to extend the phenomenologically 
interesting three-f\mbox{}lavour quark model which combines the 
chiral $U(3)_L\times U(3)_R$ NJL-type Lagrangian with the 't 
Hooft six-quark determinant (NJLH), by supplying it with f\mbox{}lavour 
mixing eight-quark interactions. The original NJLH Lagrangian gives 
a good description of the pseudoscalar nonet, especially the 
$\eta$ and $\eta'$ masses and mixing \cite{Bernard:1988}, and 
in this form the model has been widely and successfully explored 
at the mean-field level \cite{Weise:1990}-\cite{Bernard:1993}. 

This approximation was ref\mbox{}ined by works of H. Reinhardt and R. 
Alkofer \cite{Reinhardt:1988}, who used the functional integral 
method to bosonize the model. This approach hinges decisively on the
stationary phase asymptotics of the generating functional and allows
to calculate the contribution of the classical path already at lowest 
order. This lowest order result sums all tree diagrams in the 
perturbative series in powers of the coupling constant of the 't 
Hooft interaction \cite{Osipov:2005}. 

The functional treatment of the model reveals one essential problem: 
the model has actually several classical trajectories which belong to 
the interval of the functional integration, and therefore contribute to 
the integral \cite{Osipov:2005}. If one takes them into account, the 
ef\mbox{}fective potential of the theory gets unbounded from below, 
i.e., the system does not have a ground state.

We argue here that this drawback of the NJLH model can be removed.  
The eight-quark interactions added to the original Lagrangian reduce 
(under given conditions) the number of stationary phase trajectories 
to one and, as a result, the theory has a stable global minimum,
attributed to a spontaneous symmetry breakdown. It should be remarked 
that the stationary phase equations which appear in this approach are 
of cubic order and have, in general, more than one admissible
solution. We obtain inequalities for coupling constants to distinguish 
those solutions further and show that these constraints can be 
f\mbox{}inally understood as the stability criteria of the whole
system. We consider the most general eight-quark spin zero
interactions invariant under $U(3)_L\times U(3)_R$ chiral symmetry and 
assume that current quarks have realistic masses: 
$\hat{m}_u\neq\hat{m}_d\neq\hat{m}_s$. It is shown that our result is 
independent both of the specif\mbox{}ic form of eight-quark
interactions and values of current quark masses.

Let us note that only one type of the eight-quark 
interactions considered are f\mbox{}lavour mixing (the f\mbox{}irst part of the eight-quark 
Lagrangian studied here, i.e., ${\cal L}_1$ in eq.(\ref{L1})) and have 
been used previously \cite{Alkofer:1989} in a dif\mbox{}ferent 
context, namely to introduce OZI-violating ef\mbox{}fects 
\cite{Okubo:1963} in a NJL-type model with the $U_\A (1)$ anomaly 
term inspired by the works of Di Vecchia and Veneziano 
\cite{Veneziano:1980}, and independently by Rosenzweig, 
Schechter and Trahern \cite{Schechter:1980}. Recently, by describing 
the properties of nuclear matter with two-f\mbox{}lavour NJL models, 
eight-fermion interactions of the ${\cal L}_1$-type have been also
analyzed in \cite{Madland:2003}.


\section*{\centerline{\large\bf 2. The model}}

The dynamics of the model considered is determind by the Lagrangian 
density
\begin{equation}
\label{efflag}
  {\cal L}_\eff =\bar{q}(i\gamma^\mu\partial_\mu -\hat{m})q
          +{\cal L}_{\njl} + {\cal L}_{\h}
          +{\cal L}_{8q}
\end{equation}
where it is assumed that quark f\mbox{}ields have colour $(N_c=3)$ and 
f\mbox{}lavour $(N_f=3)$ indices. The current quark mass, $\hat{m}$, is a 
diagonal matrix with elements $\mbox{diag} (\hat{m}_u, \hat{m}_d, 
\hat{m}_s)$, which explicitly breaks the global chiral $SU_L(3)\times 
SU_R(3)$ symmetry of the Lagrangian.  The f\mbox{}lavour symmetry of the 
model becomes $SU(3)$, if $\hat{m}_u = \hat{m}_d = \hat{m}_s$; and one gets 
the reduced symmetries of isospin and hypercharge conservation, if 
$\hat{m}_u = \hat{m}_d\neq\hat{m}_s$. Putting $\hat{m}_u\neq\hat{m}_d
\neq\hat{m}_s$, one obtains the most general pattern of the explicit 
symmetry breakdown in the model. 

We suppose that quark verticies are ef\mbox{}fectively local, this 
being a frequently used approximation. Even in this essentially 
simplif\mbox{}ied form the Lagrangian has all basic ingredients to 
describe the dynamical symmetry breaking of the hadronic vacuum and 
f\mbox{}ind its stability condition.

The interaction Lagrangian of the NJLH model in the scalar and pseudoscalar channels is given by two terms 
\begin{eqnarray}
\label{L4q}
  {\cal L}_\njl &\!\!\! =\!\!\!& \frac{G}{2}\left[(\bar{q}
  \lambda_aq)^2+ (\bar{q}i\gamma_5\lambda_aq)^2\right], \\
\label{Ldet}
  {\cal L}_\h &\!\!\! =\!\!\! &\kappa (\mbox{det}\ \bar{q}P_Lq
                      +\mbox{det}\ \bar{q}P_Rq).
\end{eqnarray}
The f\mbox{}irst one is the $U_L(3)\times U_R(3)$ chiral symmetric 
interaction specifying the local part of the ef\mbox{}fective 
four-quark Lagrangian in channels with quantum numbers $J^P=0^+, 0^-$.
The Gell-Mann f\mbox{}lavour matrices $\lambda_a,\ a=0,1,\ldots ,8,$ 
are normalized such that $\mbox{tr} (\lambda_a \lambda_b) =2\delta_{ab}$. 
The second term represents the 't Hooft determinantal interactions 
\cite{Hooft:1978}. The matrices $P_{L,R}=(1\mp\gamma_5)/2$ are 
projectors and the determinant is over f\mbox{}lavour indices. The 
determinantal interaction breaks explicitly the axial $U_\A (1)$ 
symmetry \cite{Diakonov:1996} and Zweig's rule. 

The new feature of the model is the inclusion of $U(3)_L\times U(3)_R$ 
symmetric eight-quark forces, which we add to the standard NJLH 
Lagrangian to obtain the stable ground state. They are described by 
the term ${\cal L}_{8q}={\cal L}_1+{\cal L}_2$, where 
\begin{eqnarray}
\label{L1}    
   {\cal L}_1&\!\!\! =\!\!\! & 
   8g_1\left[ (\bar q_iP_Rq_m)(\bar q_mP_Lq_i) \right]^2, \\ 
\label{L2}   
   {\cal L}_2&\!\!\! =\!\!\!& 
   16 g_2 (\bar q_iP_Rq_m)(\bar q_mP_Lq_j) 
   (\bar q_jP_Rq_k)(\bar q_kP_Lq_i). 
\end{eqnarray}
The f\mbox{}lavour indices $i,j,\ldots =1,2,3=u,d,s$, and $g_1, g_2$ 
stand for the various symmetric eight-quark coupling strengths. The 
f\mbox{}irst term ${\cal L}_1$ coincides with the OZI-violating 
eight-quark interactions considered in \cite{Alkofer:1989}. The second 
term ${\cal L}_2$ represents interactions without violation of Zweig's 
rule. ${\cal L}_{8q}$ is the most general Lagrangian which describes 
the spin zero eight-quark interactions without derivatives. It is the 
lowest order term in number of quark f\mbox{}ields which is relevant 
to the case. We restrict our consideration to these forces, because in 
the long wavelength limit the higher dimensional operators are 
suppressed. 

Large $N_c$ arguments can be also used to justify this step
if the dimensionfull coupling constants $[G]=M^{-2}, [\kappa ]=M^{-5}, 
[g_1]=[g_2]=M^{-8}$ count at large $N_c$ as $G\sim 1/N_c$, $\kappa\sim 
1/N_c^{N_f}$, $g_1, g_2\sim 1/N_c^4$. In this case the NJL interactions 
(\ref{L4q}) dominate over ${\cal L}_\h$ and ${\cal L}_{8q} $ at large 
$N_c$, as it should be, because Zweig's rule is exact at $N_c=\infty$. 
On the other hand, with these counting rules the Lagrangians 
${\cal L}_\h$ and ${\cal L}_{8q}$ contribute at the same $N_c$
order, thus the ef\mbox{}fects coming from them are comparable and
must be considered together\footnote{Let us note that our counting for 
$g_1$ dif\mbox{}fers from the prescription of paper \cite{Alkofer:1989} 
where $g_1\sim 1/N_c^3$.}. 
      
It is clear that our considerations are also relevant if the multi-quark 
interactions create a hierarchy \cite{Simonov:2002} similar to the 
hierarchy found within the gluon field correlators \cite{Bali:2001}. 
In this case the lowest four-quark interaction forms a stable vacuum 
corresponding to spontaneously broken chiral symmetry. The higher 
multi-quark interactions in the hierarchy must not destroy this state, 
otherwise they would be as important as the lowest order terms. 
Since however the 't Hooft interaction, which is the next term in 
the hierarchy, destroys the ground state \cite{Osipov:2005}, 
one cannot truncate the tower of multi-quark interactions at this
level. The next natural candidate is the eight-quark term 
${\cal L}_{8q}$. We show that its inclusion is suf\mbox{}ficient to 
stabilize the ground state.      


\section*{\centerline{\large\bf 3. The eight-quark term at work}}

The many-fermion vertices of Lagrangian ${\cal L}_\eff$ can be 
presented in the bilinear form by introducing the functional 
unity \cite{Reinhardt:1988} in the vacuum-to-vacuum amplitude of the 
theory. The specif\mbox{}ic details of this bosonization procedure are 
given in our recent work \cite{Osipov:2005}. The new interaction
term ${\cal L}_{8q}$, which we add now to the ef\mbox{}fective quark 
Lagrangian, does not create additional problems, and the method can be 
simply extended to the present case. This is why we take as a starting 
point the corresponding functional integral already in its bosonized form 
\begin{eqnarray}
\label{genf3}
   Z&\!\!\! =\!\!\!&\int {\cal D}q{\cal D}\bar{q}
     \prod_a{\cal D}\sigma_a\prod_a{\cal D}\phi_a\ 
     \exp\left(i\int\ud^4x
     {\cal L}_q(\bar{q},q,\sigma ,\phi )\right)
     \nonumber \\
    &\!\!\!\times\!\!\!& \int\limits^{+\infty}_{-\infty}
     \prod_a{\cal D}s_a\prod_a{\cal D}p_a\
     \exp\left(i\int\ud^4x{\cal L}_r(\sigma ,\phi ,\Delta ;s,p)\right),
\end{eqnarray}  
where
\begin{eqnarray}
\label{lagr2}
  {\cal L}_q &\!\!\! =\!\!\!&
  \bar{q}(i\gamma^\mu\partial_\mu -m-\sigma - i\gamma_5\phi )q, \\
\label{Lr}
   {\cal L}_r &\!\!\! =\!\!\!& s_a (\sigma_a+\Delta_a) + p_a\phi_a
   +\frac{G}{2} \left(s_a^2+p_a^2 \right) 
   \nonumber \\
   &\!\!\! +\!\!\!& \frac{\kappa}{32}\ A_{abc}s_a
   \left(s_bs_c-3p_bp_c\right)
   + \frac{g_1}{8} \left(s_a^2+p_a^2\right)^2
   \nonumber \\
   &\!\!\! +\!\!\!& \frac{g_2}{8}\left[d_{abe}d_{cde}\left(
   s_as_bs_cs_d + 2s_as_bp_cp_d +p_ap_bp_cp_d \right)\right.
   \nonumber \\
   &\!\!\! +\!\!\!& \left. 4f_{ace}f_{bde} s_as_bp_cp_d\right]. 
\end{eqnarray}
It is worth to observe that we did not use any approximations to
obtain this result. 

Let us explain our notations. The bosonic f\mbox{}ields 
$\sigma_a$ and $\phi_a$ are the composite scalar and pseudoscalar 
nonets which will be identif\mbox{}ied later with the corresponding 
physical states. The auxiliary f\mbox{}ields $s_a$ and $p_a$ must be 
integrated out from the ef\mbox{}fective mesonic Lagrangian 
${\cal L}_r$. We assume that $\sigma =\sigma_a\lambda_a$, and so on
for all bosonic f\mbox{}ields $\sigma ,\phi ,s,p$. The quarks obtain 
their constituent masses $m=m_a\lambda_a=\mbox{diag}(m_u,m_d,m_s)$ due 
to dynamical chiral symmetry breaking in the physical vacuum state, 
$\Delta_a=m_a-\hat{m}_a$. The totally symmetric constants $A_{abc}$
are related to the f\mbox{}lavour determinant, and equal to
\begin{equation}
\label{A}
   A_{abc}=\frac{1}{3!}\epsilon_{ijk}\epsilon_{mnl}(\lambda_a)_{im}
             (\lambda_b)_{jn}(\lambda_c)_{kl}. 
\end{equation}
 
The eight-quark interactions change drastically the semi-classical 
asymptotics of the functional integral over $s_a, p_a$ in
(\ref{genf3}), as compared to the case, when $g_1, g_2=0$. To see this 
one should f\mbox{}irst f\mbox{}ind all real stationary phase 
trajectories $s_a^{st}=s_a(\sigma, \phi ),\ p_a^{st}=p_a(\sigma, 
\phi )$ given by the equations   
\begin{equation} 
\label{fdLr}
   \frac{\partial {\cal L}_r}{\partial s_a} = 0, \qquad 
   \frac{\partial {\cal L}_r}{\partial p_a} = 0. 
\end{equation}
We seek these solutions in form of expansions in the external mesonic 
f\mbox{}ields, $\sigma_a , \phi_a$, 
\begin{eqnarray}
\label{Rst}
   s_a^{st} &\!\!\! =\!\!\!& h_a + h_{ab}^{(1)}\sigma_b  
            + h_{abc}^{(1)}\sigma_b\sigma_c 
            + h_{abc}^{(2)}\phi_b\phi_c + \ldots 
            \nonumber \\
   p_a^{st} &\!\!\! =\!\!\!& h_{ab}^{(2)}\phi_b 
            + h_{abc}^{(3)}\phi_b\sigma_c + \ldots 
\end{eqnarray}
The coef\mbox{}f\mbox{}icients $h_{a\ldots}^{(i)}$ depend on the 
coupling constants $G,\kappa, g_1, g_2$ and quark masses $\Delta_a$. 
The higher index coef\mbox{}f\mbox{}icients $h_{a\ldots}^{(i)}$ are 
recurrently expressed in terms of the lower ones. The one-index 
coef\mbox{}f\mbox{}icients $h_a$ are the solutions of the following 
system of cubic equations   
\begin{equation}
\label{ha}
   \Delta_a + Gh_a + \frac{3\kappa}{32}\ A_{abc}h_bh_c
   +\frac{g_1}{2}\ h_ah_b^2 +\frac{g_2}{2}\ 
   d_{abe}d_{cde}h_bh_ch_d = 0.
\end{equation}
The trivial solution $h_a=0$, corresponds to the perturbative vacuum 
$\Delta_a=0$. There are also non-trivial ones. In accordance with the 
pattern of explicit symmetry breaking the mean f\mbox{}ield $\Delta_a$ 
can have only three non-zero components at most with indices
$a=0,3,8$. It means that in general we have a system of only 
three equations to determine $h_a\lambda_a=\mbox{diag} (h_u,h_d,h_s)$ 
\begin{equation}
\label{saddle-1}
   \left\{ \begin{array}{l}
\vspace{0.2cm}   
   Gh_u + \Delta_u +\displaystyle\frac{\kappa}{16}\ h_dh_s
   +\displaystyle\frac{g_1}{4}\ h_u(h_u^2+h_d^2+h_s^2)
   +\displaystyle\frac{g_2}{2}\ h_u^3=0, \\
\vspace{0.2cm}   
   Gh_d + \Delta_d +\displaystyle\frac{\kappa}{16}\ h_uh_s
   +\displaystyle\frac{g_1}{4}\ h_d(h_u^2+h_d^2+h_s^2)
   +\displaystyle\frac{g_2}{2}\ h_d^3=0, \\
\vspace{0.2cm}   
   Gh_s + \Delta_s +\displaystyle\frac{\kappa}{16}\ h_uh_d
   +\displaystyle\frac{g_1}{4}\ h_s(h_u^2+h_d^2+h_s^2)
   +\displaystyle\frac{g_2}{2}\ h_s^3=0. 
   \end{array} \right.
\end{equation}

Our aim now is to show that parameters can be f\mbox{}ixed in such a 
way that this system will have only one real solution. We start by 
summing the f\mbox{}irst two equations, which leads to the cubic 
equation 
\begin{eqnarray}
   && x^3+tx=b, \nonumber \\
   && t=\frac{1}{g_1+g_2}\left(
      8G + \frac{\kappa}{2}\ h_s 
      +y^2(g_1+3g_2) + 2g_1 h_s^2\right), \nonumber \\
   && b=-\frac{8(\Delta_u +\Delta_d)}{g_1+g_2},
\end{eqnarray}
where $x=h_u+h_d, \quad y=h_u-h_d$.

Note that deviations of the variable $y$ from zero are a measure of isospin breaking effects 
due to electromagentic forces, as the difference $h_u-h_d$ does not vanish for $\hat{m}_u\ne \hat{m}_d$. 
The function $t(y,h_s)$ has a minimum (if $g_1>0$ and $g_1+3g_2>0$) 
at $y=0$ and $h_s=-\kappa /(8g_1)$, thus the inequality $t>0$ always 
holds for coupling constants f\mbox{}ixed by
\begin{equation}
\label{stab}
   G > \frac{1}{g_1}\left(\frac{\kappa}{16}\right)^2.
\end{equation}
In this case the cubic equation has for any given value of $b$ just one
real root
\begin{equation}
\label{x1}
   x_{(1)} = \left(\frac{b}{2}+\sqrt{D}\right)^{\frac{1}{3}}
     + \left(\frac{b}{2}-\sqrt{D}\right)^{\frac{1}{3}}, \quad
   D=\left(\frac{t}{3}\right)^3+\left(\frac{b}{2}\right)^2.
\end{equation}
Since $b<0$ (provided that $\Delta_u +\Delta_d>0$), this function is 
negative. Its minimum is located at the point $y=0,\ h_s=-\kappa
/(8g_1)$, and the surface $x=0$ is an asymptotic one to $x_{(1)}$.

Subtracting the second equation from the f\mbox{}irst one we obtain
a quadratic equation with respect to $x$. Its solutions are given by 
\begin{equation}  
\label{x2}
   x_{(2)}=\pm \sqrt{\frac{-1}{g_1+3g_2}\left(8G-\frac{\kappa}{2}\ h_s 
     +2g_1h_s^2+y^2(g_1+g_2)+\frac{8(\Delta_u-\Delta_d)}{y} \right)}. 
\end{equation}
Since we only allow real solutions, the following inequality must hold 
\begin{equation}  
   8G-\frac{\kappa}{2}\ h_s+2g_1h_s^2+y^2(g_1+g_2)
   < \frac{8}{y}\left(\Delta_d-\Delta_u\right). 
\end{equation}
For def\mbox{}initeness, we suppose that $\Delta_d-\Delta_u>0$. This
assumption represents one of two possible alternatives. Our 
f\mbox{}inal mathematical conclusions do not depend on the choice
made. However, it is not obvious which of them should be required 
physically. Next, the function $f(h_s)=8G-\kappa h_s /2 +2g_1h_s^2>0$, 
since the minimum value $f_{min}=8G-\kappa^2/(32g_1)>0$ in the parameter 
region of (\ref{stab}). Therefore 
$y$ ranges over the half-open interval $0<y\leq y_{max}(h_s)$. The
lower bound $y=0$ is an asymptotic surface for the function
$x_{(2)}$. The upper bound $y_{max}(h_s)$ is the unique real solution 
of the equation $y^3(g_1+g_2)+yf(h_s)+8(\Delta_u-\Delta_d)=0$. It
follows that $y_{max}\propto (\Delta_d -\Delta_u)$, i.e., the 
electromagnetic forces which are responsible for the isospin symmetry 
breaking determine the length of the segment $[0,y_{max}]$, which is 
relatively small as compared with intervals determined by the strong
interaction. As a consequence a negative branch of the function 
$x_{(2)}$ grows rapidly with $y$ from $-\infty$ at $y=0$ up to $0$ at 
$y=y_{max}$.  On the contrary, functions $x_{(1)}(y)$ and $x_{(3)}(y)$ 
(see eq.(\ref{x3}) below) remain almost unchanged in the interval 
$0<y<y_{max}$, because here the strong driving forces totally 
cover electromagnetic ef\mbox{}fects.

\begin{figure}[t]
\centerline{\epsfig{file=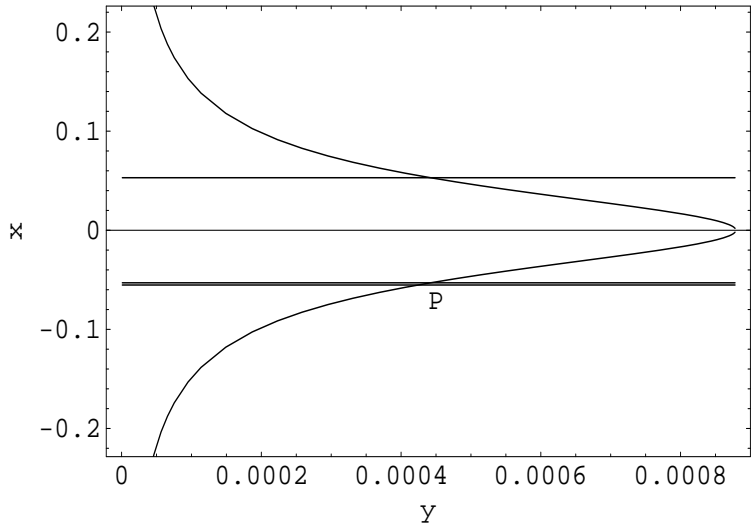,height=7.0cm,width=10cm}}
\caption{\small The label $x$ stands for the curves $x_{(i)}(h_s,y)$ 
  plotted as function of $y$ at f\mbox{}ixed $h_s=-0.03\,\mbox{GeV}^3$. 
  The bellshaped curve corresponds to the two branches of $x_{(2)}$, 
  its peak is located at $y_{max}(h_s)=8.8\cdot 10^{-4}\,\mbox{GeV}^3.$ The 
  thick line indicates the negative branch of curve $x_{(3)}$ and also 
  the curve $x_{(1)}$, which are degenerate at this scale, the upper 
  line is the positive branch of $x_{(3)}$. The two branches of
  $x_{(3)}$ meet at $y\simeq 0.14\,\mbox{GeV}^3$, outside the
  indicated range in the plot. The real solution of the system 
  (\ref{saddle-1}) is indicated by the point $P$. The corresponding 
  parameters are $G=6.1\,\mbox{GeV}^{-2},\ 
  g_1=g_2=4\cdot 10^3\,\mbox{GeV}^{-8},\ 
  \kappa=-705\,\mbox{GeV}^{-5},\
  \Delta_u=313\,\mbox{MeV},\ \Delta_d=318\,\mbox{MeV},\
  \Delta_s=345\,\mbox{MeV}$.}
\label{fig1}
\end{figure}

Let us consider now the third equation which yields
\begin{equation}
\label{x3}  
   x_{(3)}=\pm 4\ \sqrt{\frac{v(h_s,y)}{-(\kappa +8g_1h_s)}}\ , 
\end{equation}
where we have introduced the notation
\begin{equation}  
   v(h_s,y)=(g_1+2g_2)h_s^3 
   +\frac{h_s}{2}(8G+g_1y^2)
   -\frac{\kappa}{16} y^2+4\Delta_s. 
\end{equation}
The expression under the square root is positive, if conditions 
\begin{equation}
\label{in2}
   v(h_s,y)>0, \quad \kappa +8g_1h_s<0,
\end{equation}
are fulf\mbox{}illed. The alternative case does not have solutions, 
since we assume that $\Delta_s>0$ and $\kappa<0$ (phenomenological 
requirements). Inequalities (\ref{in2}) hold with $h_s$ belonging to 
the half-open interval $h_s^{min}\leq h_s<h_s^{max}$. Here 
$h_s^{max}=-\kappa /(8g_1)>0$, and $h_s^{min}<0$. The lower bound is a 
solution of the equation $v(h_s,y)=0$. This cubic equation has only
one real root which is negative for 
\begin{equation}
\label{in3}
   g_1+2g_2>0, \quad 8G +g_1y^2>0,
   \quad 4\Delta_s-\frac{\kappa}{16}\ y^2>0.
\end{equation}
Under the assumptions made above these inequalities are obviously 
fulf\mbox{}illed.

We illustrate the case in two f\mbox{}igures. The $y$-dependence is 
shown in f\mbox{}ig.1. Since $x_{(2)}$ is a monotonic function of $y$ 
in the region $x<0,\ 0<y<y_{max}$ at any f\mbox{}ixed value of $h_s$, 
the question whether the system (\ref{saddle-1}) has one or more
solutions is now reduced to a careful check of the number of 
intersections for curves $x_{(1)}$ and $x_{(3)}$ as functions of $h_s$ 
at a fixed value of $y$. Actually, for this purpose one can choose any 
value of $y$ from the interval $0<y<y_{max}$, because functions
$x_{(1)}$ and $x_{(3)}$ are almost insensitive to this value.

In f\mbox{}ig.2 we show $x_{(1)}(h_s,y)$ and $x_{(3)}(h_s,y)$ as 
functions of $h_s$, at f\mbox{}ixed $y$ given by the solution $P$ of 
f\mbox{}ig.1. It is quite easy to verify that the line
$h_s=h_s^{max}$, being the asymptote for the curve given 
by eq.(\ref{x3}), crosses the other curve (\ref{x1}) in its minimum, 
dividing it in two monotonic parts. Thus, both functions decrease 
monotonically with increasing $h_s$ in the third quadrant of the 
Cartesian coordinates system formed by the line $h_s=h_s^{max}$
and the axis of abscissas. The curves have only one intersection, which 
corresponds to an unique solution of the system (\ref{saddle-1}).

\begin{figure}[t]
\centerline{\epsfig{file=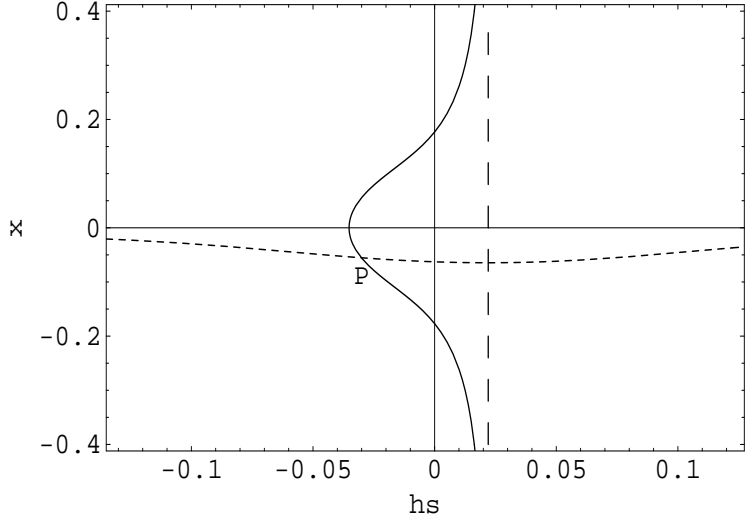,height=7.0cm,width=10cm}}
\caption{\small The curves $x_{(1)}$ of eq.(\ref{x1}) (small dashed 
         line) and $x_{(3)}$ of eq.(\ref{x3}) (solid line) are shown 
         as functions of $h_s$ for the parameter set of f\mbox{}ig.1, 
         at f\mbox{}ixed $y=4.2\cdot 10^{-4} \mbox{GeV}^3$. The 
         solution $P$ has $h_u=-0.02747,\ h_d=-0.02789,\ h_s=-0.03$ 
         in units $\mbox{GeV}^3$. The vertical dashed line corresponds 
         to $h_s=-\kappa /(8g_1 )$.} 
\label{fig2}
\end{figure}

The fact that the cubic equations (\ref{saddle-1}) have only one set
of real roots over a certain range of values of parameters $G, \kappa, 
g_1, g_2, \Delta_i$ is crucial for the ground state of the theory: it 
makes the vacuum globally stable\footnote{Let us recall that  
putting $g_1=g_2=0$, one obtains from (\ref{saddle-1}) the system of 
quadratic equations to f\mbox{}ind $h_u, h_d, h_s$. It has been shown 
in \cite{Osipov:2005} that such equations have two real solutions 
(for a physical set of parameters) in the  $SU(3)$ case and three
real solutions in the $SU(2)\times U(1)$ case. This is exactly the 
underlying reason for the vacuum instability. }. 

Unfortunately, we merely can f\mbox{}ind the solution $(h_u, h_d, h_s)$ 
numerically, apart from the simplest case with the octet f\mbox{}lavour 
symmetry, where current quarks have equal masses  
$\hat{m}_u=\hat{m}_d=\hat{m}_s$, and the system (\ref{saddle-1}) 
reduces to a cubic equation for only one variable $h_u=h_d=h_s$ 
\begin{equation}
\label{cubeq1}
   h_u^3 + \frac{\kappa}{12\lambda}\, h_u^2
         +\frac{4G}{3\lambda}\, h_u + \frac{4\Delta}{3\lambda}=0
\end{equation}
with $\lambda =g_1+(2/3)g_2$. Making the replacement 
$h_u=\bar{h}_u-\kappa /(36\lambda )$, one obtains from (\ref{cubeq1})
\begin{equation}
\label{cubeq2}
   \bar{h}_u^3 + t'\bar{h}_u = b',
\end{equation}
where 
\begin{equation}
   t'=\frac{4}{3} \left[ \frac{G}{\lambda} - 
   \left(\frac{\kappa}{24\lambda}\right)^2\right],\quad
   b'=\frac{4}{3}\left\{ 
   \frac{\kappa}{36\lambda}\left[\frac{G}{\lambda}
   -\frac{2}{3}\left(\frac{\kappa}{24\lambda}
   \right)^2\right] -\frac{\Delta}{\lambda} \right\}.
\end{equation}  
It is clear now that this cubic equation has one real root, if $t'>0$, 
i.e.,
\begin{equation}  
\label{stabcond}
   \frac{G}{\lambda}>\left(\frac{\kappa}{24\lambda}\right)^2.
\end{equation}
In this particular case the proof of existence and uniqueness of the
solution is straightforward. Let us also note that the solution found
above for the general case deviates not much from the case with octet
symmetry, i.e., we have approximately $h_u\simeq h_d\simeq h_s$.


\section*{\centerline{\large\bf 4. Ef\mbox{}fective potential}}

Since the system of equations (\ref{fdLr}) can be solved, we are able 
to obtain the semi-classical asymptotics of the integral over $s_a,
p_a$ in (\ref{genf3}). One has the following result which is valid at 
lowest order of the stationary phase approximation
\begin{eqnarray}
\label{intJisp}
     {\cal Z}[\sigma, \phi, \Delta ] &\!\!\! =\!\!\!&
     \int\limits^{+\infty}_{-\infty} \prod_a{\cal D}s_a
     \prod_a{\cal D}p_a\
     \exp\left(i\int\ud^4x{\cal L}_r(\sigma, \phi, \Delta ; s, p)\right)
     \nonumber \\
     &\!\!\!\sim\!\!\!& {\cal N}\sum_{j=1}^n\ 
     \exp\left(i\int\ud^4x{\cal L}_r(\sigma, \phi, \Delta ; 
     s_\st^{(j)} , p_\st^{(j)})\right)
     \quad (\hbar\to 0),
\end{eqnarray} 
where $n$ is the number of real solutions $(s^{\st}_a,
p^{\st}_a)^{(j)}$ of eq.(\ref{fdLr}).

The information about the vacuum state is contained in the 
ef\mbox{}fective potential of the theory. To obtain it let us consider 
the linear term in the $\sigma_a$ f\mbox{}ield. The resulting 
contribution, as it follows from eq.(\ref{intJisp}), is
\begin{equation}
\label{splo}
     {\cal Z}\sim \exp \left(i\int\ud^4x \sum_{j=1}^n 
     h_a^{(j)}\sigma_a + \ldots\right)
\end{equation}
This part of the Lagrangian is responsible for the dynamical symmetry 
breaking in the multi-quark system and taken together with the 
corresponding part from the Gaussian integration over quark 
f\mbox{}ields in eq.(\ref{genf3}) leads us to the gap equations 
(for each of quark's flavours $i=u,d,s$),
\begin{equation}
   \sum_{j=1}^n h_i^{(j)} + \frac{N_c}{2\pi^2}m_iJ_0(m_i^2) = 0,
\end{equation}
where $J_0(m_i^2)$ is the tadpole quark loop contribution with a 
high-momentum cutof\mbox{}f $\Lambda$
\begin{equation}
   J_0(m_i^2)=\Lambda^2 - m_i^2\ln\left(1+\frac{\Lambda^2}{m_i^2}
   \right).
\end{equation}

Using standard techniques \cite{Osipov:2004}, we obtain from the 
gap-equations the ef\mbox{}fective potential $U(m_i)$ as a function of 
the constituent quark masses $m_i$ which corresponds, in general, to the 
case with $n$ real roots. Here it is more convenient to use 
$(h_u, h_d, h_s)$ as independent variables, with masses $m_i$ being
determined by eqs.(\ref{saddle-1}). In particular, if the  parameters of
the model are f\mbox{}ixed in such a way that eqs.(\ref{saddle-1})
have only one real solution, the ef\mbox{}fective potential (up to an 
unessential constant, which is omitted here) is
\begin{eqnarray} 
\label{effpot1}
     U(h_u,h_d,h_s) &\!\!\! =\!\!\!& \frac{1}{16}
     \left( 4Gh_i^2  + \kappa h_uh_dh_s + \frac{3g_1}{2} 
     \left(h_i^2\right)^2 +3g_2 h_i^4\right) 
     \nonumber \\
     &\!\!\! -\!\!\!& \frac{1}{2}\left( v(m_u^2)+v(m_d^2)
     +v(m_s^2)\right),
\end{eqnarray}
where $h_i^2=h_u^2+h_d^2+h_s^2$, $h_i^4=h_u^4+h_d^4+h_s^4$, and
\begin{equation}
   v(m_i^2)=\frac{N_c}{8\pi^2} \left[ m_i^2J_0(m_i^2) 
     +\Lambda^4 \ln\left(1+\frac{m_i^2}{\Lambda^2}\right)\right].
\end{equation}

In the specif\mbox{}ic and limited case where one deals with the octet
$SU(3)$ symmetric model and $\hat{m}_i=0$ the ef\mbox{}fective potential 
$U(m)$ is an even function of $m$ for $\kappa =0$ and its plot has the 
standard form of the double well (``mexican hat'') with two symmetric 
minima, at $m=\pm m_{\mathrm{min}}$, and one local maximum, at $m=0$. 
The 't Hooft interaction ($\kappa\neq 0$) makes this curve asymmetric: 
if $\kappa <0$, the minimum located at positive values of $m$ gets 
deeper as compared with the other minimum at negative $m$, becoming 
therefore the global minimum for the whole ef\mbox{}fective
potential. It corresponds to the stable ground state of the system with 
spontaneously broken chiral symmetry. 

To appreciate the correlation found between the number of critical
points and stability let us consider the same $SU(3)$ symmetric 
model in the range with three real roots. In this case
\begin{equation}
   \sum_{j=1}^3 h_u^{(j)}=-\frac{\kappa}{12\lambda}
\end{equation}
and we find 
\begin{equation} 
\label{effpot3}
     V(m) = \frac{\kappa}{8\lambda}\, m
     - \frac{3N_c}{16\pi^2} \left[ m^2J_0(m^2) 
     +\Lambda^4 \ln\left(1+\frac{m^2}{\Lambda^2}\right)\right].
\end{equation}

As opposed to $U(m)$ the potential with three real roots, described 
by the function $V(m)$ has at most a metastable vacuum, for 
$\kappa /\lambda >0$. If $\kappa /\lambda <0$, the ef\mbox{}fective 
potential does not have extrema in the region $m>0$. In both cases
the theory related with $V(m)$ is unbounded from below and is 
physically nonsensical. 

Several special properties of the eight-quark interactions are 
expressed in these results.  
 
F\mbox{}irstly, the couplings of the considered model can always be 
chosen to fulf\mbox{}ill the inequalities required, i.e.,
\begin{equation}
\label{ineq}
   g_1>0,\quad g_1+3g_2>0,\quad Gg_1>(\kappa /16)^2
\end{equation}
and therefore the model can be simply driven to its stable regime 
where only one critical point determines the asymptotical dynamics 
of the system. The NJLH model does not have a proper mechanism for 
that.

Secondly, the eight-quark terms dominate at large values of $h_i$,
making $U(h_i)$ positive in all directions $h_i\to\pm\infty$. As a 
result, the function $U(h_i)$ is bounded from below, and exhibits 
a global ground state. 
 
Thirdly, the ef\mbox{}fective potential (\ref{effpot1}) coincides at 
$g_1, g_2=0$ with the potential obtained in the framework of the NJLH 
model by the mean-field method \cite{Hatsuda:1994}. This is probably
at the heart of the success of NJHL: although the limit $g_1, g_2\to 0$ in 
$U(h_i)$ is formally not allowed as soon as inequality (\ref{stab}) 
does not hold (instead the system is then described by the unstable 
potential $V(h_i)$), the eight-quark terms in eq.(\ref{effpot1}) are
not likely to destroy the results of the mean f\mbox{}ield approach.
Nevertheless one should expect some new noticeable ef\mbox{}fects from 
it. 


We must conclude that the eight-quark interactions play a fundamental 
role in the formation of the stable ground state for the unstable system 
described by the NJLH Lagrangian. We consider this f\mbox{}inding as
the main result of our study.


\section*{\centerline{\large\bf 5. Summary and discussion}}

Let us summarize what we have found. 

(1) An eight-quark extension of the conventional three-f\mbox{}lavour 
NJL model with the explicit $U_\A (1)$ breaking by the 't Hooft 
determinant has been suggested.  We have taken the eight quark 
interactions in its most general form for spin zero states.
Eight-quark interactions prove to be essential in stabilizing the 
vacuum of the theory: the quark model considered follows the general 
trend of spontaneous breakdown of chiral symmetry, and possesses a 
globally stable ground state, when relevant inequalities in terms of 
the coupling constants hold. 

(2) The so ensured stability of the ground state is crucial for 
applications of the model to the study of cases in which corrections 
(radiative, temperature, density, and so on ef\mbox{}fects)
may qualitatively change the structure of the theory, e.g., by turning 
minima in the ef\mbox{}fective potential into maxima. Presently the 
$U(3)_L\times U(3)_R$ chiral symmetric NJL model with the six-quark 
't Hooft interactions is frequently used for that. The eight-quark 
extension of the model considered here is needed for well-founded 
calculations in this f\mbox{}ield.  

(3) The eight-quark interactions are an additional (to the 't Hooft 
determinant) source of OZI-violating ef\mbox{}fects. They are  of the 
same order, for $g_1\sim 1/N_c^4$. It is important to take 
them into account from the phenomenological point of view: the details
of OZI-violation are still a puzzle of nonperturbative QCD 
\cite{Sapozhnikov:2003}.

\section*{Acknowledgements}
This work has been supported by grants provided by Funda\c c\~ao para
a Ci\^encia e a Tecnologia, POCTI/FNU/50336/2003 and POCI/FP/63412/2005. 
This research is part of the EU integrated infrastructure initiative 
Hadron Physics project under contract No.RII3-CT-2004-506078. A. A. 
Osipov also gratefully acknowledges the Funda\ca o Calouste Gulbenkian for 
f\mbox{}inancial support.


\end{document}